%% file: main.tex
\documentclass[10pt,conference,twocolumn]{IEEEtran}

\usepackage{cite}
\usepackage{amsmath,amsfonts}
\usepackage{physics}
\usepackage{algorithm}
\usepackage[indLines=false,commentColor=blue]{algpseudocodex}
\usepackage{graphicx}
\usepackage{textcomp}
\usepackage{xcolor}
\usepackage[hidelinks]{hyperref}
\usepackage{svg}
\usepackage{tikz}
\usetikzlibrary{quantikz2}
\usepackage{relsize}
\usepackage{moresize}
\usepackage{subcaption}
\usepackage{multirow}
\usepackage{tabularx}
\usepackage{booktabs}
\usepackage{comment}
\usepackage{balance}
\usepackage{xcolor}

\algrenewcommand\algorithmicindent{1em}%
\title{Exploring Asymmetric QEC Code Concatenation}

\author{
    \IEEEauthorblockN{%
        Sayam Sethi\IEEEauthorrefmark{1}\IEEEauthorrefmark{3}\IEEEauthorrefmark{4},
        Maxwell Poster\IEEEauthorrefmark{1}\IEEEauthorrefmark{4},
        Aditi Awasthi\IEEEauthorrefmark{1},
        Willers Yang\IEEEauthorrefmark{2},
        Joshua Viszlai\IEEEauthorrefmark{2} and
        Jonathan Mark Baker\IEEEauthorrefmark{1}%
      }
    \IEEEauthorblockA{\IEEEauthorrefmark{1}The University of Texas at Austin, \IEEEauthorrefmark{2}University of Chicago\\
    \IEEEauthorrefmark{3}\href{mailto:sayams@utexas.edu}{sayams@utexas.edu};
    \IEEEauthorrefmark{4}Denotes equal contribution}
}

\IEEEoverridecommandlockouts

\begin{document}

\maketitle

\begin{abstract}
Concatenated quantum error-correcting codes have recently gained popularity because they improve the effective distance of a joint code without requiring the discovery of new codes with desirable parameters. A common building block for such studies is the $[[4,1,2]]$ Iceberg code, the smallest error-detecting code and therefore the one with the lowest resource overhead. However, it has an asymmetric number of $X$ and $Z$ checks, and concatenating it produces skewed logical error rates (LERs) between the $X$ and $Z$ observables. This asymmetry can be mitigated by Clifford-deforming the individual blocks of the concatenation, but the space of possible Hadamard deformations grows doubly exponentially with the number of concatenation levels. We formalize this deformation space and investigate it by proposing six concrete deformation strategies, which we compare under a code capacity (with and without bias) noise model using a concatenated
maximum-likelihood soft-information decoder.
Among the strategies we study, one one of them achieves near-identical $X$ and $Z$ LERs in a logical-memory experiment while simultaneously reducing the total LER (the sum of the $X$ and $Z$ error rates).
\end{abstract}


\section{Introduction}
Concatenated Quantum Error Correcting codes (QECC) have recently gained popularity~\cite{gidney_yoked_2023,litinski_blocklet_2025,xu_batched_2025,nakai_subsystem_2026} due to their ability to improve effective code distances of the joint code, without the need to discover and study newer codes with desired code parameters $[[n,k,d]]$. One of the most common codes used for concatenation studies is the $[[4, 1, 2]]$ code, also known as the Iceberg code. It is the smallest error detecting code and thus requires minimal resource overhead (in qubits, gates, etc) to realize its implementation. Additionally, a universal gate set for the Iceberg code has been well studied~\cite{self_protecting_2024,ginsberg_quantum_2025}, with various demonstrations  using this code for early fault-tolerance applications~\cite{dasu_computing_2026,rines_demonstration_2025}. However, the Iceberg code has an asymmetric number of $X$ and $Z$ checks ($2\times X$ checks and $1\times Z$ check, or vice versa) and concatenating the Iceberg code leads to skewed logical error rates between the $X$ and the $Z$ observables. Since concatenation requires the use of soft-information decoding, double the number of checks provides more soft-information to the decoder at the next level, despite the unconcatenated Iceberg code having equal distances in the $X$ and $Z$ basis. We can address this asymmetry in the soft-information by deforming the individual blocks of concatenation, however, the space of possible choices for Hadamard deformation grows as $\mathcal{O}(\sim2.5^{4^l})$, where $l$ is the number of levels of concatenation.
\par We investigate this space of concatenation by proposing $6$ possible strategies and report the trade-offs between them under a phenomenological noise model and a maximum-likelihood decoder. 
Out of the strategies we explore, we observe that the deformation strategy which performs more mixing between the $X$ and $Z$ checks achieves near-identical logical error rates in the $X$ and $Z$ basis when simulated in a logical memory experiment, and also reduces the total logical error rate (sum of $X$ and $Z$ error rates). We describe the deformation strategies considered in this work further in Section~\ref{sec:deformations}.
This work makes the following contributions:
\begin{itemize}
    \item We formalize the space of Hadamard deformations of a concatenated Iceberg code obtained by Hadamard conjugation, and show that it scales doubly exponentially with the concatenation level $l$ (Section~\ref{sec:deformations}).
    \item We propose six concrete reproducible deformation strategies that span natural axes of variation---per-logical qubit, per-level, and per-block (Section~\ref{sec:strategies}).
    \item We evaluate all six strategies under unbiased and biased noise across one to four levels of concatenation, and identify \texttt{Alternate\_Mixed\_within\_Level} as achieving the most symmetric and lowest total LER (Section~\ref{sec:results}).
\end{itemize}

\begin{figure}
    \centering
    \includegraphics[width=0.65\linewidth]{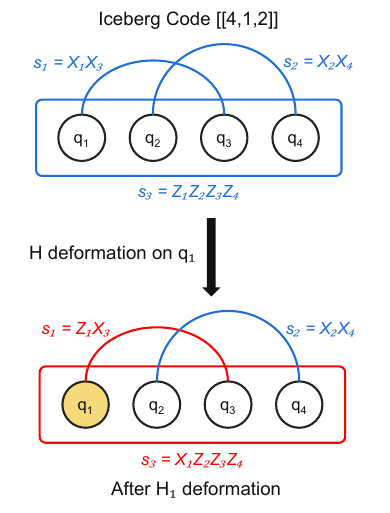}
    \caption{The $[[4,1,2]]$ Iceberg code and the effect of a Clifford deformation. \textbf{Top:} the code on four data qubits, with two weight-2 $X$ checks ($s_1,s_2$) and one weight-4 $Z$ check ($s_3$). \textbf{Bottom:} conjugating $q_1$ by a Hadamard swaps $X\leftrightarrow Z$ on that qubit, turning $s_1$ and $s_3$ into mixed-type checks}
    \label{fig:intro}
\end{figure}

\input{figures/strategies}

\section{Background}
\subsection{Iceberg Code}
The Iceberg code is a family of error-\emph{detecting} codes defined by the parameters $[[k+2,\,k,\,2]]$. Since its distance is $2$, it can detect (but not correct) any single-qubit error, which enables error mitigation by post-selection. The smallest instance is the $[[4,1,2]]$ code, constructed from the $[[4,2,2]]$ code by fixing one of its two logical qubits as a gauge qubit.

The stabilizers for the $[[4,1,2]]$ Iceberg code are given by $\{X_1X_3,\, X_2X_4,\, Z_1Z_2Z_3Z_4\}$ and its logical operators $X_L, Z_L$ are given by $X_1X_2$ and $Z_1Z_3$, respectively. More generally, the checks of the $[[k+2,k,2]]$ family have the form
\begin{equation*}
    S_X = X_t X_b \prod_i X_i, \qquad
    S_Z = Z_t Z_b \prod_i Z_i,
\end{equation*}
with logical operators $\overline{X}_i = X_t X_i$ and
$\overline{Z}_i = Z_b Z_i$, where $t$ and $b$ denote the two ``boundary'' qubits. A key structural feature (and the source of the problem we study) is that the $X$-type and $Z$-type checks are not symmetric: the code has different number of $X$ and $Z$ stabilizers, so it protects the two Pauli observables unequally.

\begin{figure*}
    \centering
    \includegraphics{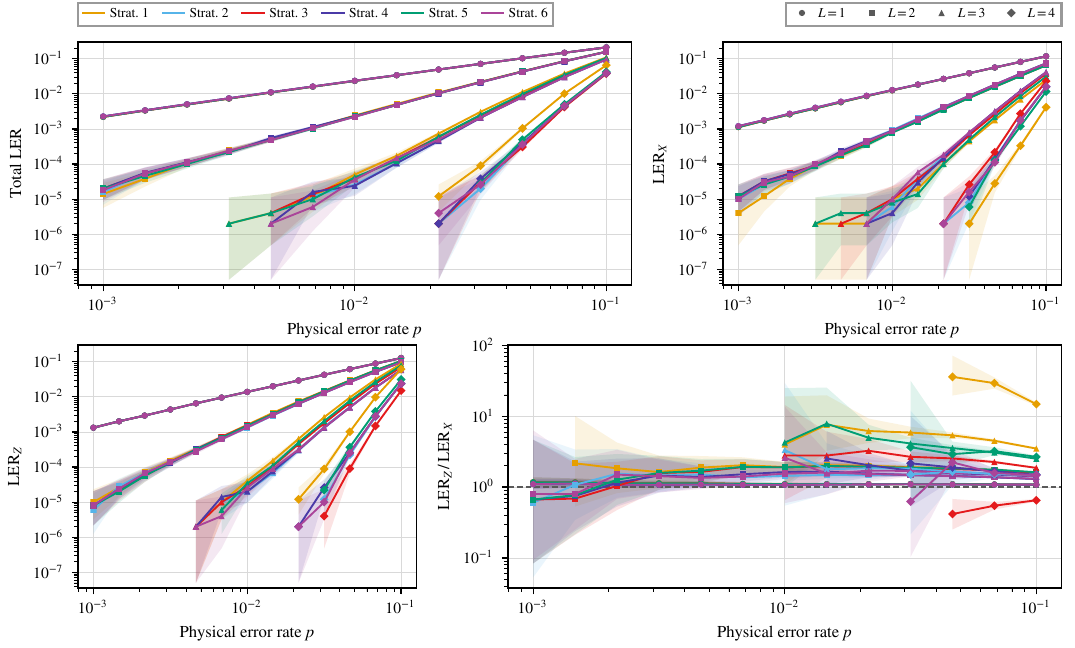}
    \caption{LER under unbiased noise across one to four levels of concatenation.}\label{fig:ler-plots}
\end{figure*}


\subsection{Clifford Deformations}
A stabilizer code can be deformed by conjugating its stabilizers and logical observables by a Clifford $\mathcal{C}$. While the stabilizer group changes, the code remains equivalent to the original code in the sense that:
\begin{itemize}
    \item The encoded logical subspace is preserved
    \item The code distance remains unchanged, despite the effective distance being varied
    \item The topological properties are maintained in the case of single-qubit deformations
\end{itemize}

Recent work has explored random CDSCs~\cite{Dua_2024}, where deformations are randomly assigned from $\{I, H, HS\}$ with probabilities $(\Pi_I, \Pi_{XZ}, \Pi_{YZ})$. Remarkably, the $(0.25, 0.5)$ random CDSC was found to outperform both XZZX and XY codes at moderate bias $\eta \sim 100$, with typical random realizations beating the best-known translationally invariant codes. This suggests that carefully chosen deformation patterns can achieve performance superior to simpler symmetric constructions.

We consider the restricted set of deformations, $\mathcal{D}_1$, obtained by conjugating the Iceberg code with Hadamard gates on an arbitrary subset of the data qubits. Applying a Hadamard $H$ on a qubit has the effect of swapping the $X$ and $Z$ operators on that qubit, as shown in Fig.~\ref{fig:intro}. This can be further generalized to arbitrary single-qubit Clifford deformations, however, it comes with its own set of challenges. Firstly, we will need to modify the decoder to support non-CSS codes. Since we use the maximum-likelihood decoder, this will blow up the exponent in the complexity term of the decoder from $2$ to $4$, thereby, squaring the time taken. Furthermore, the search space for possible Clifford deformations blows up from $2$ possible deformations per-qubit to $3! = 6$ possible deformations per qubit, which further blows up when considering concatenation.

\subsection{Code Concatenation}
Concatenation of two QEC codes $C_1$ (inner) and $C_2$ (outer) proceeds as follows:
\begin{itemize}
    \item \textbf{Encoding:} each logical qubit of the inner code $C_1$
    is used as a physical qubit of the outer code $C_2$.
    \item \textbf{Logical operators:} the logical operators of the joint
    code are the products of the $C_1$ logical operators over the
    physical qubits involved in the corresponding $C_2$ operator.
    \item \textbf{Decoding:} we first decode $C_1$ via
    maximum-likelihood (ML) decoding, then propagate the soft
    information (the likelihood of each correction) from $C_1$ up to the
    ML decoder for $C_2$.
\end{itemize}
In this work, each level of the concatenation is itself an Iceberg block that can be deformed independently. The combinatorial freedom of the deformation choice is what gives rise to the search problem we analyze next.

\subsection{Depolarizing Noise Model}
Under the depolarizing model, each qubit $i$ is subjected independently to a random Pauli error $e_i$ with probability $p_i(e_i)$, where $e_i \in \{X,Y,Z\}$ and $p_i(I) = 1 - \sum_{e_i \neq I} p_i(e_i)$:
\begin{equation*}
    \mathcal{E}(\rho) =
    \sum_{i \in [n],\, e_i \in \{I,X,Y,Z\}} p_i(e_i)\, e_i\, \rho\, e_i.
\end{equation*}
Because the channel applies only Pauli operators, errors and their commutation relations with the stabilizers are easy to track. Any error $E_k$ decomposes over the Pauli basis as $E_k = \sum_{e_i \in \{I,X,Y,Z\}} \alpha_{e_i} e_i$, and when it acts on a state $\overline{\rho}$ encoded by a code with stabilizers $S = \langle S_1, \ldots, S_{n-k}\rangle$, measuring syndrome $S_i$ gives
\begin{equation*}
    S_i E_k \overline{\rho}\, E_k^{\dagger}
    = \sum_{e_i, e_i' \in \{I,X,Y,Z\}}
      \alpha_{e_i}\alpha^{*}_{e_i'}\, \lambda_i\,
      e_i\, \overline{\rho}\, e_i',
\end{equation*}
where $\lambda_i = +1$ if $e_i$ commutes with $S_i$ and $\lambda_i = -1$ if it anticommutes. Hence syndrome $S_i$ flips if and only if $E_k$ has a component that anticommutes with $S_i$; for a single Pauli error this probability is either $0$ or $1$.

\subsection{Maximum-Likelihood Decoding}\label{sec:mldecode}
Given the depolarizing model, the log-likelihood that an error on qubit $q_i$ flips a syndrome is
\begin{equation*}
    w_i(s_i) = \log\!\left(\frac{p_i(s_i)}{1 - p_i(s_i)}\right)
\end{equation*}
\begin{equation*}
    \text{where }p_i(s_i) =
    \begin{cases}
        0, & s_i = I\\
        \sum_{e_i \neq s_i} p_i(e_i), & \text{otherwise}
    \end{cases}
\end{equation*}
and $s_i$ is the Pauli parity measured on the $i$-th qubit. For every observed subset of syndrome flips $S \subseteq \{s_1, \ldots, s_{n-k}\}$, the most likely correction is
\begin{equation*}
    \mathcal{C}(S) =
    \operatorname*{arg\,max}_{\substack{
        C \subseteq \{P_1, \ldots, P_n\} \\
        \bigoplus_{P_i \in C} H^{\top}(P_i) = S}}
    \ \sum_{P_i \in C} w_i(P_i),
\end{equation*}
where $H^{\top}(P_i)$ is the row of stabilizers measuring the
$P \in \{X,Y,Z\}$ parity of qubit $i$. Computing $\mathcal{C}(S)$ is $\#\mathrm{P}$-hard for general stabilizer
codes~\cite{iyer2013decoding} (the classical analogue is $\mathrm{NP}$-complete), so exact maximum-likelihood decoding costs $\Theta(\mathrm{poly}(n,k)\,4^n)$ and is intractable in general. For concatenated codes we apply it level by level: the inner code is decoded
first, and its soft information (the correction likelihoods) is propagated to the maximum-likelihood decoder of the next-outer code. This is repeated until we reach the root, or the topmost  level, decoder/code.
 
\begin{figure*}
    \centering
    \includegraphics{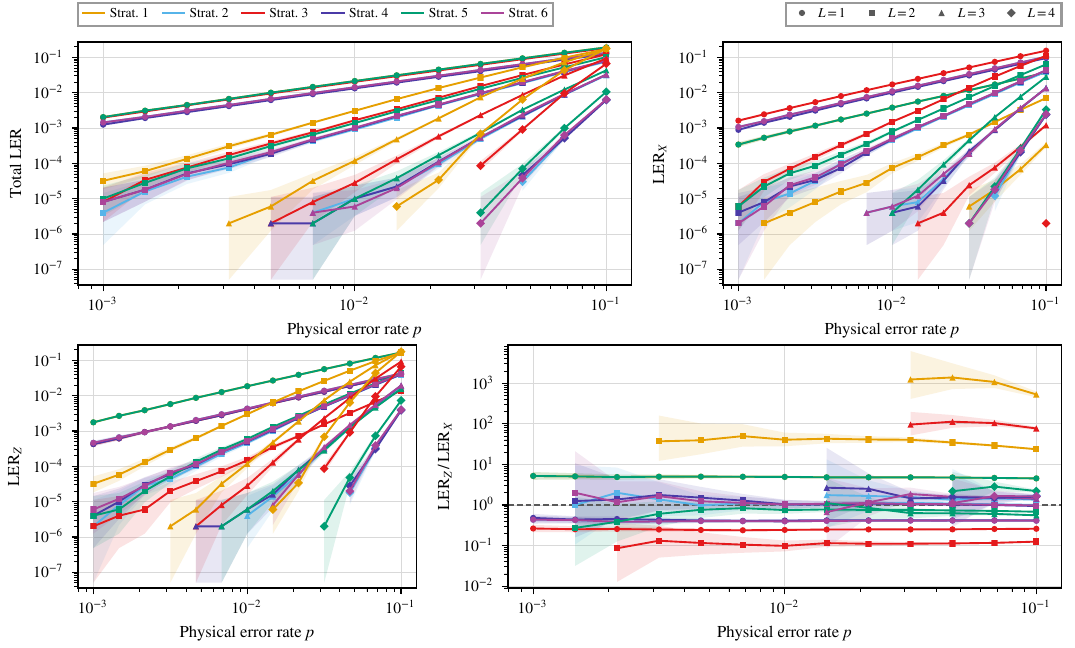}
    \caption{LER under $X$-biased noise (with $X$ errors being $8\times$ more likely than $Z$ and $Y$ errors) across one to four levels of concatenation.}
    \label{fig:xbias}
\end{figure*}

\section{Deformations of the Iceberg Code}
\label{sec:deformations}
A single Iceberg block admits $|\mathcal{D}_1| = c_1 = 2^4 = 16$ Hadamard deformations (Fig.~\ref{fig:intro} shows one such deformation). Under concatenation, this freedom compounds: the number of possible deformation choices scales via the recurrence \[c_l = c_{l-1}^4\cdot c_1 = 2^{\frac{4\cdot(4^l - 1)}{3}},\]
for $l$ levels of concatenation.
That is, adding every level lets us choose $c_1$ possible deformations for the new level, and adds $4$ codes from level $l-1$, each of which have $c_{l-1}$ deformation possibilities. 
As a result, for a given noise landscape and a desired concatenation level $l$, the problem of searching for the optimal deformation strategy scales doubly exponentially with $l$ even under a restricted deformation search space and a uniform error landscape.

\subsection{Deformation Strategies}
\label{sec:strategies}
To investigate the effect of different deformation strategies on the logical error rate of the concatenated code, we define $6$ different strategies (see Fig~\ref{fig:strategies}) and compare their effects. We pick these strategies since they allow mixing of the $X$ and $Z$ syndrome checks to varying degrees:
\begin{enumerate}
    \item\label{item:1} \texttt{No\_Deformation}: This is the standard Iceberg code.
    \item \texttt{Mixed\_Deformation}: We use the deformation $H_1 I_2 H_3 I_4$, which results in one $X$ check with weight $2$, one $Z$ check with weight $2$, and one combined $X, Z$ check with weight $4$.
    \item\label{item:3} \texttt{Alternate\_XZ\_by\_Level}: We use no deformation at every odd level, and $H_1 H_2 H_3 H_4$ deformation at every even level of concatenation.
    \item\label{item:4} \texttt{Alternate\_Mixed\_by\_Level}: We alternate between $H_1 I_2 H_3 I_4$ and $I_1 H_2 I_3 H_4$ at every even and odd level.
    \item\label{item:5} \texttt{Alternate\_XZ\_within\_Level}: Within each level, the blocks alternate between $H_1H_2H_3H_4$ and no deformation.
    \item\label{item:6} \texttt{Alternate\_Mixed\_within\_Level}: Within each level, the blocks alternate between $H_1 I_2 H_3 I_4$ and $I_1 H_2 I_3 H_4$.
\end{enumerate}
 
For the phenomenological model used in our experiments, we take $p_x = p_z = p$ in the unbiased case. To model noise biased along a Pauli $P$, we sample $P$ with probability $0.8\,p_i$ at physical error rate $p_i$ on qubit $i$, and each of the remaining two Paulis with probability $0.1\,p_i$.
 
\begin{figure*}
    \centering
    \includegraphics{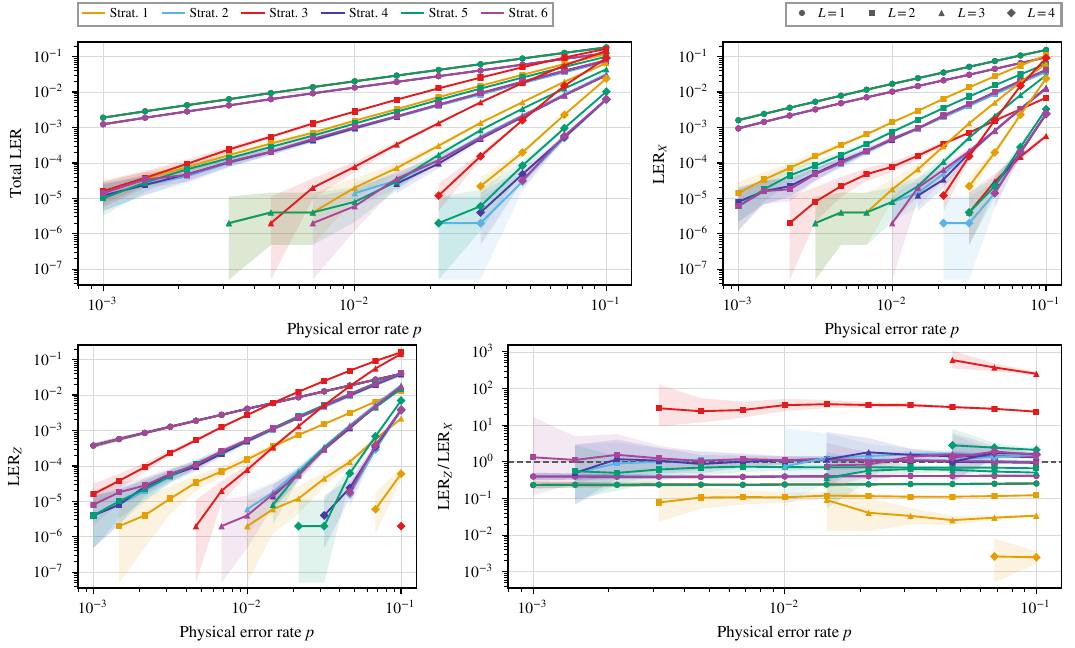}
    \caption{LER under $Z$-biased noise (with $Z$ errors being $8\times$ more likely than $X$ and $Y$ errors) across one to four levels of concatenation}
    \label{fig:zbias}
\end{figure*}

\section{Results}
\label{sec:results}
\subsection{Performance of Deformation Strategies}
We demonstrate the performance of different deformation strategies by plotting the total LER, as well as the LER for individual $X$ and $Z$ memory experiments in Figure~\ref{fig:ler-plots}. These experiments were performed under a depolarizing code capacity noise model, wherein we model the probability of a \emph{physical} bit-flip, phase-flip, and both bit- and phase-flip as $p_X = p_Z = p_Y = p/3$, where $p$ is the physical error rate. We ran $500,000$ Monte-Carlo shots of the maximum-likelihood decoder as described in Section~\ref{sec:mldecode} on the different deformation strategies. We also plot the ratio between the $Z$ and $X$ LERs. We observe that Strategy~\ref{item:1} can exhibit a $Z$ error rate up to $60\times$ larger than the $X$ error rate at $4$ levels of concatenation. However, other strategies demonstrate much lower variation in logical error rates across observables. Between Strategy~\ref{item:5} and Strategy~\ref{item:6}, the \texttt{Alternate\_Mixed\_within\_Level} strategy performs the best in both overall LER and individual memory experiments. These initial results further motivate the need to explore this problem in detail and design a heuristic that can output a robust deformation for a given concatenation level and input error model.

\subsection{Biased Noise}
We repeat the experiment under $X$-biased and $Z$-biased noise, sampling the biased Pauli with probability $0.8\,p_i$ and each of the other two Paulis with probability $0.1\,p_i$. Since the $[[4,1,2]]$ code is a CSS code and the deformations we consider only swap the $X$ and $Z$ operators, the $Y$-biased case reduces to the unbiased case: a $Y$ error splits into an $X$ and a $Z$ error with equal weight. We therefore report only the $X$- and $Z$-biased regimes, shown in Figures~\ref{fig:xbias} and~\ref{fig:zbias}.
 
Under a biased channel, the benefit of a deformation that redistributes protection between the two Pauli sectors becomes more pronounced: strategies that rebalance the checks lower the LER along the dominant (biased) axis, whereas the undeformed code concentrates failures in the sector it already protects least. Consistent with the unbiased case, \texttt{Alternate\_Mixed\_within\_Level} remains the most balanced across the settings tested.

\section{Conclusion}
We studied concatenated $[[4, 1, 2]]$ Iceberg codes and compared six block-deformation strategies for balancing logical error rates across X and Z observables. Under a phenomenological noise model with maximum-likelihood decoding, we find that deformation choice strongly affects both symmetry and total logical error rate. Among the strategies tested, Strategy 6, \texttt{Alternate\_Mixed\_within\_Level}, achieves near-symmetric logical error rates while also reducing the combined X and Z logical error rate, making it a promising approach for concatenated Iceberg-code logical memories, though further exploration must be done to design a robust deformation for arbitrary concatenation level and error model.

\section*{Acknowledgements}
The authors acknowledge the Texas Advanced Computing Center (TACC) at The University of Texas at Austin for providing computational resources that have contributed to the research results reported in this paper. This work was funded in part by the Texas Quantum Institute (TQI).

\bibliographystyle{IEEEtranS}
\bibliography{references}

\end{document}

%% file: figures/strategies.tex
\begin{figure*}
  \centering
\begin{tikzpicture}[
  scale=0.6,
  transform shape,
  rect/.style={
    draw,
    rectangle,
    minimum width=0.9cm,
    minimum height=0.32cm,
    inner sep=2pt,
    fill=white,
    text=black
  },
  edgelbl/.style={
    font=\small,
    inner sep=1pt,
    fill=white,
    text=black
  },
  contdot/.style={
    yshift=-6pt,
    inner sep=0pt
  },
  panel/.style={draw, rounded corners=2pt},
  strat/.style={
    draw,
    rounded corners=2pt,
    fill=white,
    text=black,
    font=\scriptsize\bfseries,
    align=center,
    inner sep=2pt,
    text width=4.4cm
  }
]
 
\begin{scope}[shift={(-10,0)}]
  \draw[panel] (-4.8,-1.0) rectangle (4.8,3.2);
  \node[strat] at (0,2.9) {1. No Deformation};
 
  \coordinate (p1r) at (0,2.25);
  \coordinate (p1a) at (-3.2,1.25);
  \coordinate (p1b) at (-1.1,1.25);
  \coordinate (p1c) at ( 1.1,1.25);
  \coordinate (p1d) at ( 3.2,1.25);
 
  \coordinate (p1a1) at (-4.3,0.05);
  \coordinate (p1a2) at (-3.6,0.05);
  \coordinate (p1a3) at (-2.9,0.05);
  \coordinate (p1a4) at (-2.2,0.05);
 
  \coordinate (p1b1) at (-1.9,0.05);
  \coordinate (p1b2) at (-1.2,0.05);
  \coordinate (p1b3) at (-0.5,0.05);
  \coordinate (p1b4) at ( 0.2,0.05);
 
  \draw (p1r)--(p1a) node[edgelbl, midway, above left]  {$I$};
  \draw (p1r)--(p1b) node[edgelbl, midway, left]        {$I$};
  \draw (p1r)--(p1c) node[edgelbl, midway, right]       {$I$};
  \draw (p1r)--(p1d) node[edgelbl, midway, above right] {$I$};
 
  \draw (p1a)--(p1a1) node[edgelbl, midway, above left]  {$I$};
  \draw (p1a)--(p1a2) node[edgelbl, midway, left]        {$I$};
  \draw (p1a)--(p1a3) node[edgelbl, midway, right]       {$I$};
  \draw (p1a)--(p1a4) node[edgelbl, midway, above right] {$I$};
 
  \draw (p1b)--(p1b1) node[edgelbl, midway, above left]  {$I$};
  \draw (p1b)--(p1b2) node[edgelbl, midway, left]        {$I$};
  \draw (p1b)--(p1b3) node[edgelbl, midway, right]       {$I$};
  \draw (p1b)--(p1b4) node[edgelbl, midway, above right] {$I$};
 
  \node[rect] at (p1r) {$L_i$};
  \node[rect] at (p1a) {$L_{i-1}$};
  \node[rect] at (p1b) {$L_{i-1}$};
  \node[rect] at (p1c) {$L_{i-1}$};
  \node[rect] at (p1d) {$L_{i-1}$};
 
  \node[contdot] at (p1a1) {$\vdots$};
  \node[contdot] at (p1a2) {$\vdots$};
  \node[contdot] at (p1a3) {$\vdots$};
  \node[contdot] at (p1a4) {$\vdots$};
 
  \node[contdot] at (p1b1) {$\vdots$};
  \node[contdot] at (p1b2) {$\vdots$};
  \node[contdot] at (p1b3) {$\vdots$};
  \node[contdot] at (p1b4) {$\vdots$};
 
  \node at ( 1.1,0.55) {$\vdots$};
  \node at ( 3.2,0.55) {$\vdots$};
\end{scope}
 
\begin{scope}[shift={(0,0)}]
  \draw[panel] (-4.8,-1.0) rectangle (4.8,3.2);
  \node[strat] at (0,2.9) {2. Mixed Deformation};
 
  \coordinate (p2r) at (0,2.25);
  \coordinate (p2a) at (-3.2,1.25);
  \coordinate (p2b) at (-1.1,1.25);
  \coordinate (p2c) at ( 1.1,1.25);
  \coordinate (p2d) at ( 3.2,1.25);
 
  \coordinate (p2a1) at (-4.3,0.05);
  \coordinate (p2a2) at (-3.6,0.05);
  \coordinate (p2a3) at (-2.9,0.05);
  \coordinate (p2a4) at (-2.2,0.05);
 
  \coordinate (p2b1) at (-1.9,0.05);
  \coordinate (p2b2) at (-1.2,0.05);
  \coordinate (p2b3) at (-0.5,0.05);
  \coordinate (p2b4) at ( 0.2,0.05);
 
  \draw (p2r)--(p2a) node[edgelbl, midway, above left]  {$H$};
  \draw (p2r)--(p2b) node[edgelbl, midway, left]        {$I$};
  \draw (p2r)--(p2c) node[edgelbl, midway, right]       {$H$};
  \draw (p2r)--(p2d) node[edgelbl, midway, above right] {$I$};
 
  \draw (p2a)--(p2a1) node[edgelbl, midway, above left]  {$H$};
  \draw (p2a)--(p2a2) node[edgelbl, midway, left]        {$I$};
  \draw (p2a)--(p2a3) node[edgelbl, midway, right]       {$H$};
  \draw (p2a)--(p2a4) node[edgelbl, midway, above right] {$I$};
 
  \draw (p2b)--(p2b1) node[edgelbl, midway, above left]  {$H$};
  \draw (p2b)--(p2b2) node[edgelbl, midway, left]        {$I$};
  \draw (p2b)--(p2b3) node[edgelbl, midway, right]       {$H$};
  \draw (p2b)--(p2b4) node[edgelbl, midway, above right] {$I$};
 
  \node[rect] at (p2r) {$L_i$};
  \node[rect] at (p2a) {$L_{i-1}$};
  \node[rect] at (p2b) {$L_{i-1}$};
  \node[rect] at (p2c) {$L_{i-1}$};
  \node[rect] at (p2d) {$L_{i-1}$};
 
  \node[contdot] at (p2a1) {$\vdots$};
  \node[contdot] at (p2a2) {$\vdots$};
  \node[contdot] at (p2a3) {$\vdots$};
  \node[contdot] at (p2a4) {$\vdots$};
 
  \node[contdot] at (p2b1) {$\vdots$};
  \node[contdot] at (p2b2) {$\vdots$};
  \node[contdot] at (p2b3) {$\vdots$};
  \node[contdot] at (p2b4) {$\vdots$};
 
  \node at ( 1.1,0.55) {$\vdots$};
  \node at ( 3.2,0.55) {$\vdots$};
\end{scope}
 
\begin{scope}[shift={(10,0)}]
  \draw[panel] (-4.8,-1.0) rectangle (4.8,3.2);
  \node[strat] at (0,2.9) {3. Alternate XZ by Level};
 
  \coordinate (p3r) at (0,2.25);
  \coordinate (p3a) at (-3.2,1.25);
  \coordinate (p3b) at (-1.1,1.25);
  \coordinate (p3c) at ( 1.1,1.25);
  \coordinate (p3d) at ( 3.2,1.25);
 
  \coordinate (p3a1) at (-4.3,0.05);
  \coordinate (p3a2) at (-3.6,0.05);
  \coordinate (p3a3) at (-2.9,0.05);
  \coordinate (p3a4) at (-2.2,0.05);
 
  \coordinate (p3b1) at (-1.9,0.05);
  \coordinate (p3b2) at (-1.2,0.05);
  \coordinate (p3b3) at (-0.5,0.05);
  \coordinate (p3b4) at ( 0.2,0.05);
 
  \draw (p3r)--(p3a) node[edgelbl, midway, above left]  {$H$};
  \draw (p3r)--(p3b) node[edgelbl, midway, left]        {$H$};
  \draw (p3r)--(p3c) node[edgelbl, midway, right]       {$H$};
  \draw (p3r)--(p3d) node[edgelbl, midway, above right] {$H$};
 
  \draw (p3a)--(p3a1) node[edgelbl, midway, above left]  {$I$};
  \draw (p3a)--(p3a2) node[edgelbl, midway, left]        {$I$};
  \draw (p3a)--(p3a3) node[edgelbl, midway, right]       {$I$};
  \draw (p3a)--(p3a4) node[edgelbl, midway, above right] {$I$};
 
  \draw (p3b)--(p3b1) node[edgelbl, midway, above left]  {$I$};
  \draw (p3b)--(p3b2) node[edgelbl, midway, left]        {$I$};
  \draw (p3b)--(p3b3) node[edgelbl, midway, right]       {$I$};
  \draw (p3b)--(p3b4) node[edgelbl, midway, above right] {$I$};
 
  \node[rect] at (p3r) {$L_i$};
  \node[rect] at (p3a) {$L_{i-1}$};
  \node[rect] at (p3b) {$L_{i-1}$};
  \node[rect] at (p3c) {$L_{i-1}$};
  \node[rect] at (p3d) {$L_{i-1}$};
 
  \node[contdot] at (p3a1) {$\vdots$};
  \node[contdot] at (p3a2) {$\vdots$};
  \node[contdot] at (p3a3) {$\vdots$};
  \node[contdot] at (p3a4) {$\vdots$};
 
  \node[contdot] at (p3b1) {$\vdots$};
  \node[contdot] at (p3b2) {$\vdots$};
  \node[contdot] at (p3b3) {$\vdots$};
  \node[contdot] at (p3b4) {$\vdots$};
 
  \node at ( 1.1,0.55) {$\vdots$};
  \node at ( 3.2,0.55) {$\vdots$};
\end{scope}
 
\begin{scope}[shift={(-10,-4.9)}]
  \draw[panel] (-4.8,-1.0) rectangle (4.8,3.2);
  \node[strat] at (0,2.9) {4. Alternate Mixed by Level};
 
  \coordinate (p4r) at (0,2.25);
  \coordinate (p4a) at (-3.2,1.25);
  \coordinate (p4b) at (-1.1,1.25);
  \coordinate (p4c) at ( 1.1,1.25);
  \coordinate (p4d) at ( 3.2,1.25);
 
  \coordinate (p4a1) at (-4.3,0.05);
  \coordinate (p4a2) at (-3.6,0.05);
  \coordinate (p4a3) at (-2.9,0.05);
  \coordinate (p4a4) at (-2.2,0.05);
 
  \coordinate (p4b1) at (-1.9,0.05);
  \coordinate (p4b2) at (-1.2,0.05);
  \coordinate (p4b3) at (-0.5,0.05);
  \coordinate (p4b4) at ( 0.2,0.05);
 
  \draw (p4r)--(p4a) node[edgelbl, midway, above left]  {$I$};
  \draw (p4r)--(p4b) node[edgelbl, midway, left]        {$H$};
  \draw (p4r)--(p4c) node[edgelbl, midway, right]       {$I$};
  \draw (p4r)--(p4d) node[edgelbl, midway, above right] {$H$};
 
  \draw (p4a)--(p4a1) node[edgelbl, midway, above left]  {$H$};
  \draw (p4a)--(p4a2) node[edgelbl, midway, left]        {$I$};
  \draw (p4a)--(p4a3) node[edgelbl, midway, right]       {$H$};
  \draw (p4a)--(p4a4) node[edgelbl, midway, above right] {$I$};
 
  \draw (p4b)--(p4b1) node[edgelbl, midway, above left]  {$H$};
  \draw (p4b)--(p4b2) node[edgelbl, midway, left]        {$I$};
  \draw (p4b)--(p4b3) node[edgelbl, midway, right]       {$H$};
  \draw (p4b)--(p4b4) node[edgelbl, midway, above right] {$I$};
 
  \node[rect] at (p4r) {$L_i$};
  \node[rect] at (p4a) {$L_{i-1}$};
  \node[rect] at (p4b) {$L_{i-1}$};
  \node[rect] at (p4c) {$L_{i-1}$};
  \node[rect] at (p4d) {$L_{i-1}$};
 
  \node[contdot] at (p4a1) {$\vdots$};
  \node[contdot] at (p4a2) {$\vdots$};
  \node[contdot] at (p4a3) {$\vdots$};
  \node[contdot] at (p4a4) {$\vdots$};
 
  \node[contdot] at (p4b1) {$\vdots$};
  \node[contdot] at (p4b2) {$\vdots$};
  \node[contdot] at (p4b3) {$\vdots$};
  \node[contdot] at (p4b4) {$\vdots$};
 
  \node at ( 1.1,0.55) {$\vdots$};
  \node at ( 3.2,0.55) {$\vdots$};
\end{scope}
 
\begin{scope}[shift={(0,-4.9)}]
  \draw[panel] (-4.8,-1.0) rectangle (4.8,3.2);
  \node[strat] at (0,2.9) {5. Alternate XZ within Level};
 
  \coordinate (p5r) at (0,2.25);
  \coordinate (p5a) at (-3.2,1.25);
  \coordinate (p5b) at (-1.1,1.25);
  \coordinate (p5c) at ( 1.1,1.25);
  \coordinate (p5d) at ( 3.2,1.25);
 
  \coordinate (p5a1) at (-4.3,0.05);
  \coordinate (p5a2) at (-3.6,0.05);
  \coordinate (p5a3) at (-2.9,0.05);
  \coordinate (p5a4) at (-2.2,0.05);
 
  \coordinate (p5b1) at (-1.9,0.05);
  \coordinate (p5b2) at (-1.2,0.05);
  \coordinate (p5b3) at (-0.5,0.05);
  \coordinate (p5b4) at ( 0.2,0.05);
 
  \draw (p5r)--(p5a) node[edgelbl, midway, above left]  {$I$};
  \draw (p5r)--(p5b) node[edgelbl, midway, left]        {$I$};
  \draw (p5r)--(p5c) node[edgelbl, midway, right]       {$I$};
  \draw (p5r)--(p5d) node[edgelbl, midway, above right] {$I$};
 
  \draw (p5a)--(p5a1) node[edgelbl, midway, above left]  {$I$};
  \draw (p5a)--(p5a2) node[edgelbl, midway, left]        {$I$};
  \draw (p5a)--(p5a3) node[edgelbl, midway, right]       {$I$};
  \draw (p5a)--(p5a4) node[edgelbl, midway, above right] {$I$};
 
  \draw (p5b)--(p5b1) node[edgelbl, midway, above left]  {$H$};
  \draw (p5b)--(p5b2) node[edgelbl, midway, left]        {$H$};
  \draw (p5b)--(p5b3) node[edgelbl, midway, right]       {$H$};
  \draw (p5b)--(p5b4) node[edgelbl, midway, above right] {$H$};
 
  \node[rect] at (p5r) {$L_i$};
  \node[rect] at (p5a) {$L_{i-1}$};
  \node[rect] at (p5b) {$L_{i-1}$};
  \node[rect] at (p5c) {$L_{i-1}$};
  \node[rect] at (p5d) {$L_{i-1}$};
 
  \node[contdot] at (p5a1) {$\vdots$};
  \node[contdot] at (p5a2) {$\vdots$};
  \node[contdot] at (p5a3) {$\vdots$};
  \node[contdot] at (p5a4) {$\vdots$};
 
  \node[contdot] at (p5b1) {$\vdots$};
  \node[contdot] at (p5b2) {$\vdots$};
  \node[contdot] at (p5b3) {$\vdots$};
  \node[contdot] at (p5b4) {$\vdots$};
 
  \node at ( 1.1,0.55) {$\vdots$};
  \node at ( 3.2,0.55) {$\vdots$};
\end{scope}
 
\begin{scope}[shift={(10,-4.9)}]
  \draw[panel] (-4.8,-1.0) rectangle (4.8,3.2);
  \node[strat] at (0,2.9) {6. Alternate Mixed within Level};
 
  \coordinate (p6r) at (0,2.25);
  \coordinate (p6a) at (-3.2,1.25);
  \coordinate (p6b) at (-1.1,1.25);
  \coordinate (p6c) at ( 1.1,1.25);
  \coordinate (p6d) at ( 3.2,1.25);
 
  \coordinate (p6a1) at (-4.3,0.05);
  \coordinate (p6a2) at (-3.6,0.05);
  \coordinate (p6a3) at (-2.9,0.05);
  \coordinate (p6a4) at (-2.2,0.05);
 
  \coordinate (p6b1) at (-1.9,0.05);
  \coordinate (p6b2) at (-1.2,0.05);
  \coordinate (p6b3) at (-0.5,0.05);
  \coordinate (p6b4) at ( 0.2,0.05);
 
  \draw (p6r)--(p6a) node[edgelbl, midway, above left]  {$H$};
  \draw (p6r)--(p6b) node[edgelbl, midway, left]        {$I$};
  \draw (p6r)--(p6c) node[edgelbl, midway, right]       {$H$};
  \draw (p6r)--(p6d) node[edgelbl, midway, above right] {$I$};
 
  \draw (p6a)--(p6a1) node[edgelbl, midway, above left]  {$H$};
  \draw (p6a)--(p6a2) node[edgelbl, midway, left]        {$I$};
  \draw (p6a)--(p6a3) node[edgelbl, midway, right]       {$H$};
  \draw (p6a)--(p6a4) node[edgelbl, midway, above right] {$I$};
 
  \draw (p6b)--(p6b1) node[edgelbl, midway, above left]  {$I$};
  \draw (p6b)--(p6b2) node[edgelbl, midway, left]        {$H$};
  \draw (p6b)--(p6b3) node[edgelbl, midway, right]       {$I$};
  \draw (p6b)--(p6b4) node[edgelbl, midway, above right] {$H$};
 
  \node[rect] at (p6r) {$L_i$};
  \node[rect] at (p6a) {$L_{i-1}$};
  \node[rect] at (p6b) {$L_{i-1}$};
  \node[rect] at (p6c) {$L_{i-1}$};
  \node[rect] at (p6d) {$L_{i-1}$};
 
  \node[contdot] at (p6a1) {$\vdots$};
  \node[contdot] at (p6a2) {$\vdots$};
  \node[contdot] at (p6a3) {$\vdots$};
  \node[contdot] at (p6a4) {$\vdots$};
 
  \node[contdot] at (p6b1) {$\vdots$};
  \node[contdot] at (p6b2) {$\vdots$};
  \node[contdot] at (p6b3) {$\vdots$};
  \node[contdot] at (p6b4) {$\vdots$};
 
  \node at ( 1.1,0.55) {$\vdots$};
  \node at ( 3.2,0.55) {$\vdots$};
\end{scope}
 
\end{tikzpicture}  \caption{The six deformation strategies of Section~III-A. Each node is
  an Iceberg block at some concatenation level $L_i$; edge labels ($I$ or
  $H$) give the per-qubit Hadamard deformation applied to that block.}
  \label{fig:strategies}
\end{figure*}